RESEARCH

# Supporting novel biomedical research via multilayer collaboration networks

Konstantin Kuzmin[1], Xiaoyan Lu[1†], Partha Sarathi Mukherjee[1†], Juntao Zhuang[1†], Chris Gaiteri[2] and Boleslaw K Szymanski[1,3*]

*Correspondence: szymab@rpi.edu
[1]Network Science and Technology Center, Rensselaer Polytechnic Institute (RPI), 110 Eighth Street, Troy, NY 12180, USA
Full list of author information is available at the end of the article
†Equal contributor

## Abstract

The value of research containing novel combinations of molecules can be seen in many innovative and award-winning research programs. Despite calls to use innovative approaches to address common diseases, an increasing majority of research funding goes toward "safe" incremental research. Counteracting this trend by nurturing novel and potentially transformative scientific research is challenging, it must be supported in competition with established research programs. Therefore, we propose a tool that helps to resolve the tension between safe/fundable research vs. high-risk/potentially transformational research. It does this by identifying hidden overlapping interest around novel molecular research topics. Specifically, it identifies paths of molecular interactions that connect research topics and hypotheses that would not typically be associated, as the basis for scientific collaboration. Because these collaborations are related to the scientists' present trajectory, they are low risk and can be initiated rapidly. Unlike most incremental steps, these collaborations have the potential for leaps in understanding, as they reposition research for novel disease applications. We demonstrate the use of this tool to identify scientists who could contribute to understanding the cellular role of genes with novel associations with Alzheimer's disease, which have not been thoroughly characterized, in part due to the funding emphasis on established research.

## Introduction

Public calls for return on investment in biological research, such as the Cancer Moonshot [1] and the National Alzheimer's Plan Act that targets a preventative drug by 2025 [2], do not seek incremental scientific advances. Instead they call for transformative insights that will substantially improve patient care. Big data resources in biology may be one path to creating such insights, as seen in efforts to extract actionable research directions, such as European programs on organizing large-scale biological data [3] and collaborative endeavors across National Institutes of Health (NIH), like the Big Data to Knowledge trans-NIH initiative [4] or scientific community attempts to create large-scale metabolic models [5]. However, in the face of unprecedented data sources and public calls for transformative scientific research that makes use of these resources, the expert consensus is that the field of biology increasingly favors "safe" research that does not challenge the status quo of the field [6].

According to Smalheiser et al. [7], there are two extreme cases of how collaboration is established. One is a passive approach when one side, a supplier, of the relationship assumes a “vendor model” by providing only a minimal set of well-defined resources to the receiver who is typically the initiator of the collaboration. The other extreme case is an active model where two parties are fully engaged in the collaboration, carry equal responsibility, and receive equal credit for the work. There is also a wide range of possibilities between those extremes which can be potentially very productive but are very hard to initiate due to uncertainty associated with the need to agree upon many essential details. As a proposed solution, [7] introduces a set of guidelines which describe several possible engagement levels (the minimal level and a number of higher levels) that can be used by a supplier and a receiver to negotiate the terms of the collaboration. Our goal is also to support potential collaborations that can emerge from the middle area between two extremes.

The origin of overly cautious approaches in biology is likely rooted in incentives for research. The 23% drop in NIH funding since 2003 leads to fear that any negative result from a high-risk project would be catastrophic for grant applications, which pushes young scientist to work on “safe” projects [8]. Publications and grant applications now require more substantial experimental support [9], leading to a rich-get-richer cycle of support for established ideas and investigators. Indeed, there is a steadily increasing average age of R01 Research Project Grant recipients older than 65 outnumber those under 35, whose share of RO1’s has shrunken from 18% to 3% [10]. The upshot of these trends is that the vast majority of research funding occurs within the current paradigm, with little incentive to push for transformative results [11].

Proposals have been made to revise funding incentives in a way that encourages more innovation in biomedical research, by distributing funding more broadly [12] or by randomly selecting the best grants among top applications when the peer review process cannot make an accurate determination [13, 14]. Even if action is taken at a high level to implement such suggestions, the effects will take years or even full generations to become visible. Therefore, we propose a tool that can help to immediately reverse the trend towards disruptive science, but does not require “high-risk” efforts by young scientists. We do this by mining the structure of multilayer molecular and authorship networks in search for rational innovative partnerships, which have been shown to generate high-quality scientific findings [15]. The current relationship of publications to molecular networks is that publications generally pertain to “popular molecules” and rarely connect to less studied ones [16]. A more efficient way to explore biochemical relationships entails moving away from popular topics and exploring additional subjects. Accordingly, award-winning scientists show a preference for exploring emerging topics [17] and novel relationships between them [18]. We utilize molecular networks to promote innovative, unbiased science, while minimizing career risk; we identify and connect researchers whose topics of study are “nearby” in molecular networks. Essentially, when molecules $A$ and $B$ interact biophysically, we suggest that researchers of molecule $A$ and $B$ should interact scientifically to explore their related interests.

By mirroring molecular organization in science, we decrease historical bias, all while appealing to pre-standing interests, which makes collaboration less costly.

For instance, one scientist may have negative findings related to a molecule in the context of cancer, while those results can be useful to another scientist who studies interacting molecules in schizophrenia. The links between their research, which share no overlapping keywords, can only be found through the structure of molecular networks, which connect the molecules they study. These collaboration recommendations not only make use of big data (in the form of molecular networks), but are resistant to historical bias and can be updated as new or specialized molecular data become available. In short, by following paths in molecular data, we can begin to construct rational scientific communities, as we alert researchers to the hidden potential of their existing research.

To identify innovative collaborations, we merge molecular interaction networks with authorship information related to specific molecules, to create a multilayer network structure. We can then mine this multilayer network for path density between researchers to predict collaboration potential. In theory, any molecular network can be used to link researchers, and indeed such resources are evolving rapidly, with increasing accuracy, genome-wide coverage, and disease relevance [19, 20]. Different types or context of molecular networks may be most relevant to particular researchers; for instance, researchers who work with *drosophila* may wish to use interactions identified in data from that species to predict their ideal collaborators. As new interaction data become available, these adjustments to the molecular interaction component of the multilayer network will warp the distance between researchers and highlight new potential collaborations.

The remainder of this paper is organized as follows. A survey of the relevant research on multilayer networks and related fields is provided in section Related Work. The section Synergy Landscapes Architecture describes the main design and architectural features of the Synergy Landscapes project, explains our choice of the source of data and the graph storage solution, and provides details on the multilayer network organization and network analysis features that we implemented. That section also defines the two main approaches to ranking the results of queries that we designed, implemented, and compared in this study. In section Performance Evaluation we present the experimental results of our two recommendation methods and compare their performance both in terms of the actual recommendations produced by each method and their running times. Finally, Conclusion and Future Work summarizes our contribution and provides some final remarks along with an overview of our vision of the Synergy Landscapes project's future.

## Related Work

One of the key concepts of the Synergy Landscapes [21] project is to establish new collaborative links between different types of entities (molecules, authors, publications, etc.) Therefore, it is natural to combine those elements into a single data structure that should provide efficient means of storing, manipulating, and mining this information. While methods of the classic graph theory can be easily applied to the data primitives we have to model (e.g., molecules, authors, and publications can be represented as nodes, whereas edges can correspond to the relationships between the entities), combining elements of different kinds with possibly multiple types of relations between them requires a more complex data structure. In the following paragraphs we provide an overview of the existing approaches to overcoming the traditional limitations of the homogeneous monoplex networks and justify the choice we made for our project.

Another essential property of our system is the relevance of the collaborations that we identify or, more broadly, the quality of predictions made by our method. A natural way to express the relevancy of the query result is to rank the output. In the last part of this section, we discuss several existing ranking methods. This discussion should provide sufficient background to understand the approaches implemented in our solution which are described later in section Synergy Landscapes Architecture.

### Multilayer Networks

The idea of combining several different but related datasets into a single multilayer network is widely used in complex systems. De Domenico et al. [22] define multilayer networks as networks which contain entities with different sets of neighbors in each layer. The applications of multilayer networks are mostly found in sociology and social information systems. A comprehensive review by Boccaletti et al. [23] contains a detailed description of the properties and structural and dynamic organization of networks that represent different relationships as layers. Such networks have shown utility in economics, technical systems, ecology, biology and psychology. We include molecular interaction networks as a novel layer in Synergy Landscapes. These networks originate from many experimental sources and model organisms. In many omics analyses it is now standard to project results into these networks structures, to identify the overall functional role of the results or additional related molecules. Many free and commercial online tools are available for this purpose (e.g., [24] and [25]). At the same time, methodologically related studies of co-authorship and human social networks have emphasized the relevance of network structure in determining patterns of collaboration [26].

### Collaboration Networks

With the focus in recent years in researching social networks, collaboration networks among scientists have also been explored. The earliest work on this field by Newman et al. in 2001 [27] defines these networks. In these networks, a scientist is represented by a node. Two scientist nodes are joined via an edge only if they have been co-authors in a publication. Note that these edges are unweighted. Such networks can be used to explore social connections among scientists. These networks have been studied and the various network metrics calculated in [27] and [28]. These measures include means and distributions of the number of edges, clustering coefficient, average distances between scientists in a network, centrality measures like closeness and betweenness centrality.

Recent work by Bian et al. [29] goes beyond such traditional metrics. The networks themselves are slightly different — the edges are weighted based on the number of collaborative grants awarded to the relevant pair of scientists, instead of co-authorships. Multi-year grants are counted for every fiscal year. On these enhanced networks, the "leaders", or the most influential scientists are identified by various centrality measures and rank aggregation techniques. Furthermore, new collaborations are suggested using the Random Walk with Restart (RWR) algorithm. However, this research does not take into account connections between scientists who might be working on related topics but who might not have collaborated.

Guimerà et al. [30] explored how teams of collaborators are formed and the effect of the team's structure on its performance. According to their model, teams consist of newcomers and incumbents. The distribution of edges which correspond to different interactions between these two types of team members determines the overall team profile, like the level of innovation or diversity. The evolution of the network is characterized by a phase transition during which a large connected cluster is formed from many small clusters. This large connected cluster corresponds to a wide network of social and professional interactions which span across institutional boundaries. It is also shown that teams which publish in high-impact journals (it can be used as a measure of the performance) tend to have a larger relative size of the giant component. In our work, the goal is to suggest such potential collaborations which would contribute to this phase transition and increase the performance of teams.

### Heterogeneous Information Networks

In recent years, data mining in Heterogeneous Information Networks (HINs) has gained popularity. We can colloquially define an HIN as a network with nodes and edges of different types. In [31] Gong et al. give a formal mathematical definition and describe a Social-Attribute Network (SAN) which is an example of a heterogeneous network. HINs are widely used to model and study different types of networks in various fields, like social sciences, biology and medicine, and transportation, as well as across the fields (e.g., scientific collaboration networks). The fact that heterogeneous networks include different types of entities and relationships in many cases significantly simplifies the process of mapping the properties of objects being studied to the attributes of network entities, as compared to homogeneous networks. For example, in the dblp computer science bibliography [32] database one node can represent either a publication or an author. Publications and authors are connected with relationships, such as "co-author", "cite" and "cited-by", and "publish" and "published-by". Multilayer networks and HINs, though have different terminologies, can be essentially treated as networks with multiple types of nodes, while HINs highlights the different types of relationships among the nodes.

The problems including link prediction, recommendation, clustering, entity matching, and ranking have been studied in the context of HINs in the last several years. One of the earliest work, [33], proposed co-ranking of authors and publications in academic collaboration networks. It extends the original random walk algorithm for ranking entities by considering the probability of a surfer jumping between the co-authorship network and the citation network so that authors and publications are ranked simultaneously.

In order to extract semantic meanings of the data, meta-path [31], which is a sequence of edge types, is proposed for extracting the heterogeneous features for a pair of nodes. The starting time of collaboration between two researchers is predicted according to the number of instances of different meta-paths they have in [31].

### Ranking Methods

In order to recommend users with the proper profiles of collaborators, we apply ranking algorithms to the co-authorship networks to locate the researchers who are working on the molecules related to the molecule of interests. Among various ranking algorithms, PageRank [34] has been shown to correctly order items across a wide variety of applications. It uses the random walk model where at each step of the simulation a surfer jumps from one node in the network to another neighboring node with a certain probability. After a sufficient number of jumping steps, the probability of the surfer staying at a particular node represents the importance of this node. To avoid trapping the surfer in nodes without any outgoing edges, when such a node is encountered, the surfer will jump to a random node with certain probability at each step after arrival. The publication of the original PageRank algorithm resulted in numerous papers published to extend the original method, e.g., Tong et al. [35] proposed to use community structures of the network in order to speed up the computation of PageRank. In order to provide personalized recommendations, a sub-networks can be constructed dynamically and serve as input of the PageRank algorithm.

Although the PageRank algorithm manages to provide a robust and reliable ranking order, it does not guarantee that the computation will finish quickly. Since users send queries to the server expecting to receive a response in a few seconds, PageRank algorithm can not be used. For these reasons, we consider the matrix factorization algorithm which is often used as one of the alternatives to the PageRank-based recommendation. It factorizes the adjacency matrix of a large network into two matrices in which the row or column represents the attribute vector of one item. The product of attributes vectors of two items can be interpreted as a measure of similarity between them and thus can be used for ranking recommendations. The benefit of this approach is that the factorization algorithm can be executed in advance and the obtained attribute vectors can be stored in the databases. Thereafter, given an item in the network, the items most similar to it can be precomputed using the inner product of attribute vectors resulting in fast recommendations.

## Synergy Landscapes Architecture

The architectural design of Synergy Landscapes follows the modular organization which ensures a high level of independence of individual components, like the user interface and data storage and processing. Each module is developed in accordance with an explicitly defined interface which communicates with the rest of the system using standard well-defined protocols. As a result, implementation of every module can be changed without affecting any other modules or the system as a whole. This provides good scalability and extensibility.

One of the foundations of the Synergy Landscapes project is the idea of combining domain-specific heterogeneous data with possibly multiple types of relationships between the entities into a single multilayer network and then providing a way of mining and ranking these data to discover non-trivial relationships between data elements. It turns out that this essential feature of the Synergy Landscapes design makes it quite general and applicable to many domains beyond its use in biology.

In order to highlight this distinction between the general conceptual architecture that can be applied across multiple fields and specific implementations of the framework in a particular domain, we will use the name Synergy Landscapes for the former and individual proper names for the latter. In this paper we describe the version of Synergy Landscapes for use in biology. Since this implementation can be thought of as a tool which provides *hints* about collaboration based on *molecular* networks, we call it MoleClue. Another way to look at MoleClue is to treat it as a practical application of the Synergy Landscapes architecture to support novel biological research.

## General System Design and Organization

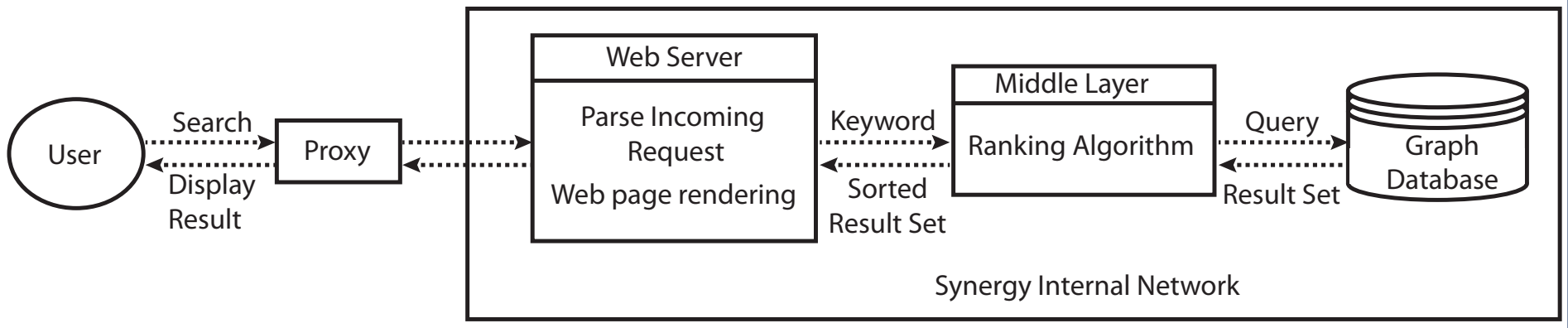

**Figure 1 Architecture of the Synergy Landscapes solution.** The modular Synergy Landscapes system is comprised of three main tiers: presentation (Web-based user interface and Web page rendering), data model (Neo4j graph database), and processing and control (data acquisition and maintenance, query processing, and ranking of the results).

Our Synergy network shown in Figure 1 consists of the proxy, web server, middle layer, and graph database.

The proxy is the entry point for user to access our network. It redirects user's search request to web server and also hosts all the static web pages of our website. The web server extracts the keyword from user's request and sends it to middle layer. After getting the result data from middle layer, it dynamically generates the web page of result and returns it to user. The middle layer composes the query for database, receives the initial result and runs our ranking algorithm on the result set.

There are several advantages of introducing a proxy to our network. Firstly, it improves security by blocking user's direct access to the internal network of the MoleClue application. Secondly, it can improve the performance of web server by buffering all incoming requests. Thirdly, it provides scalability to our network: in future when facing large number of requests we can simply deploy multiple identical web servers and middle layer behind the proxy without further modifying the system.

The middle layer is the connecting layer between the Web interface and the graph database. Based on the request from the Web layer, it connects to the database, executes the search query and fetches the results. It contains the logic for all the result processing, including the implementations for the ranking methods to sort the results.

## Data Storage

The data storage solution is an essential component of the Synergy Landscapes framework as it is central to almost any operation performed in the system. For example, data received from the external publication database should be stored locally to provide fast and convenient access. Searches requested by the users are performed using queries which should be able to efficiently extract and process information from the local database. Since the data that Synergy Landscapes is designed to handle can be naturally represented as a graph, some kind of a noSQL graph database would be a good choice for our data storage needs.

A number of studies [36], [37], [38] indicate that for the purposes of storing data that can be represented as graphs or networks graph databases offer important advantages over the traditional relational database model. We selected a popular Neo4j graph database as the core component of our data storage subsystem.

Neo4J offers a convenient programming model [39] which can be used to efficiently integrate this graph database into custom solutions. It also offers several query methods [40] (both native and RESTful) which allow developers to select the right balance between expressiveness of the syntax and runtime efficiency. Neo4j is available in two editions, one of which, the Community edition, can be used free of charge.

### Multilayer Network Organization

In order to be able to execute user queries, it is first necessary to setup a multilayer network and populate it with data. The network is built incrementally, layer by layer. In the beginning, we create a molecular layer. It contains nodes which correspond to distinct molecules and edges that connect molecules known to be associated with each other in some context. For each molecule, its name and a list of aliases is retained. One of the goals of our synergistic search process is to infer additional molecular connectivity information that is not part of the initial network of molecules. While having molecular data is required for the basic use cases of our solution, there are additional layers which can provide extended functionality and enrich our multilayer network with supplemental data but are not strictly necessary. For instance, if data on the association between molecules and diseases, molecules and grants, or researchers and their joint grants is added, it would bring greater flexibility in our ability to generate subsequent layers and enable us to perform more sophisticated user queries.

The second group of layers in our multilayer network is built using publication data. Currently, our MoleClue application uses PubMed [41] database as a source of publication data. However, the underlying Synergy Landscapes architecture imposes no limitations on the number of sources. Other publication providers can be integrated into the system to populate additional publication layers in the multilayer network.

In order to be able to associate incoming publication records with relevant molecules, the process of populating the publication layer is guided by analyzing the features and metadata of publications for specific keywords, like names of molecules. For instance, if the name of a certain molecule is found in the abstract of a publication or its list of keywords, then an edge will be created between the molecule and the publication. Thus, the publication layer of the network consists of nodes which represent publications and edges which connect publications that are known to be associated with each other in some context (e.g., which mention the same molecule in a list of keywords or the text of the abstract).

At this point, there are no edges between nodes which are located in the publication layer. Indeed, such relations can not be deduced from the source publication or molecular data alone. They will be determined after creating other layers and establishing associations between different parts of the network. Since the publication layer is created based on the search terms provided by the molecular layer, the publication–molecule cross-layer emerges naturally. Figure 2 shows a sample multilayer network with three layers (molecules, publications, and authors).

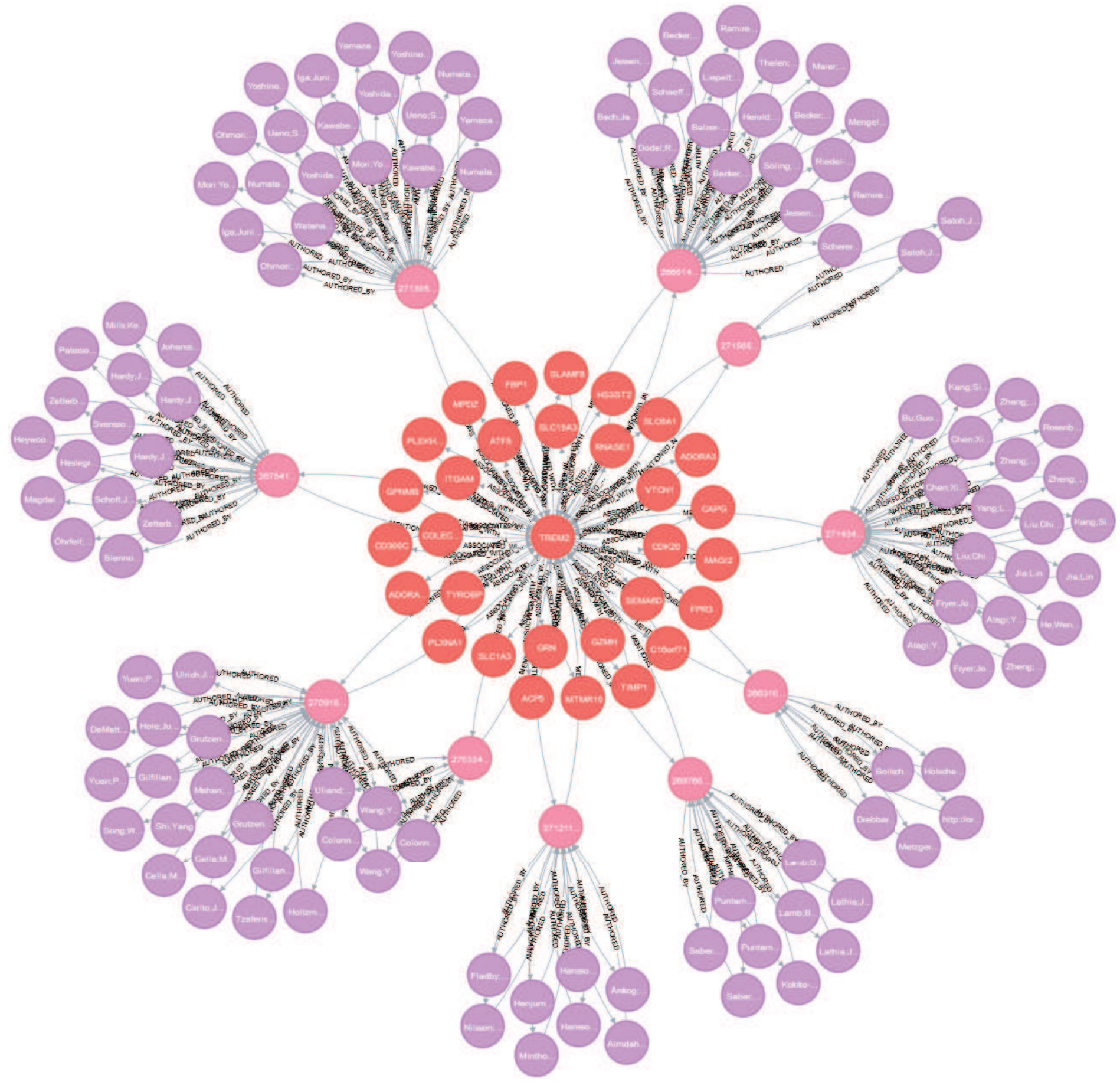

**Figure 2 TREM2 molecule and its partial neighborhood: related molecules, publications, and authors.** This sample of the MoleClue heterogeneous multilayer network shows the partial neighborhood of the TREM2 protein coding gene (some nodes and edges have been removed to reduce the clutter). Two molecules are connected with an edge if they are known to occur together in some biological context (e.g., participation in the same regulatory function). The weight on an edge corresponds to the strength of the association. An edge between a molecule and a publication exists when the publication mentions the molecule (e.g., in the title, abstract, or the list of keywords). An edge between a publication and an author expresses the authorship relation. All edges are directional but since all underlying relationships are symmetrical there is a pair of opposite directed edges between any two connected nodes.

In a publication–molecule cross-layer, an edge connects a certain molecule to the publications which are known to refer to this molecule. Such cross-layers represent layers consisting entirely of edges. Moreover, instead of connecting nodes of a single underlying node set, the edges in cross-layers go “vertically” across any two different layers, effectively “stitching” them together. Therefore, cross-layers are fundamental entities in the multilayer network since they facilitate “vertical” connectivity between layers and allow network analysis tools to traverse the whole stack of layers rather than being trapped in any single one of them. In a wider context, a cross-layer with two corresponding node sets can be regarded as a separate bipartite network linking two different entities (publications and molecules, authors and diseases, etc.)

Currently, our MoleClue application has three layers implemented: molecules, publications, and authors. As the project is developed further, more layers will be added. In its present form, our multilayer network is primarily built on the collaboration relationship between authors who published their findings related to particular molecules. However, the Synergy Landscapes architecture is not limited just to collaboration-based multilayer networks. Nodes belonging to a certain layer can be connected to nodes from other layers using any type of relationship and even multiple types of relationships.

For instance, having a disease layer with nodes corresponding to diseases and edges connecting diseases with molecules which are known to be associated with them will allow users to execute additional types of queries. One example is searching for authors working on molecule $Mol$ who also worked on diseases $d_i$ and $d_j$ and another is finding diseases that were studied by researchers who considered molecule $Mol$ in their publications. The Synergy architecture also enables more complex queries. For example, one can start with some molecule $Mol_i$ and find all diseases with which $Mol_i$ has been associated in past publications. Then it would be possible to find if some other molecules $Mol_j$ and $Mol_k$ have ever been studied with those diseases and if so what authors and publications were involved. Finally, it can be determined if a pair of molecules $Mol_i$ and $Mol_j$ is associated with different diseases and who were the experts who described those reactions in their publications.

The basic molecular network is extensible as new layers, such as those based on medications and their relationships with molecules and diseases, can be created easily. For example, in addition to the publication layer, our network can contain data on grants. This would make it possible to search for potential collaborators in writing grant proposals or to track the performance of research conducted under a particular grant as it leads to new publications.

### Network Analysis and Mining

The key use case of Synergy is discovering connections between two entities not identified by traditional overlapping interests. One such use case focuses on finding connections between a molecule, which serves as the search term and connected authors. While authors publishing on this search term can be found by querying our 3-layered network, it does not fully exploit the power of our network. The objective of Synergy is to perform a more sophisticated search, i.e., find the authors who have published on molecules one hop away from the searched molecule. This is done in the following way.

First, all the molecules one hop away from the search term are found. From this point onwards, these molecules are referred to as *related molecules*. For every related molecule, the authors who have published on them are found. When this process is completed, the results for every related molecule are merged. For example, let's assume that one related molecule is $Mol_1$ and author $Auth_1$ has published $x$ papers on it and there is another related molecule $Mol_2$ and the author $Auth_1$ has published $y$ papers on it. After merging, the results would note the author $Auth_1$'s contributions for both molecules - $Auth_1$-$Mol_1$:$x$,$Mol_2$:$y$. The sum of $x$ and $y$ is the number of publications by an author $Auth_1$ on the related molecules of the search term, denoted by $n_{PC}$. Along with this sum, the total number of publications of $Auth_1$ on all molecules, not only the related ones, are also noted. It is denoted by $n_{TOTAL}$.

The contributions of each author in the results are thus noted. However, simply obtaining a list of authors is not sufficient. A ranking mechanism is required to discern which authors are the leaders in the neighborhood of the searched molecule. We selected, implemented, and tested three ranking methods described in the following paragraphs.

The first ranking method is based on the hypergeometric test. The intuition behind this test is as follows. Given a population size and the number of successes in the population, what is the probability that a sample of a particular size will have $k$ or more successes? In this scenario, the population size is the total number of publications in the database and the number of successes in the population is the total number of publications on all the related molecules of the search term. Every sample indicates an author, the sample size, $n_{TOTAL}$, and the number of successes in the sample, $n_{PC}$, defining the value of parameter $k$ in the test.

While the hypergeometric ranking method is very powerful, an author who has many publications (close to or as high as $n_{PC}$) and mostly on one related molecule will rise in the rankings. Yet, it would be beneficial to have authors who have published on many of the related molecules, to find who "covers" the neighborhood of the search term better. This consideration led to our second publication count based on ranking approach. Here, we sort the list of names in descending order, based on the number of related molecules on which the authors have published. To break the ties between two people with the same number of related molecules in their publications, we sort them in descending order, based on their values of $n_{PC}$.

Since no normalization is performed by this ranking algorithm, it is useful for finding the most prolific authors who "cover" the neighborhood of the search term well and who publish extensively on the related molecules. However, this promotes older authors who have been publishing for longer periods of time over younger ones who have started publishing more recently. To address this, we introduce a third measure using the normalized publication count, $r_{PC}$, defined in Equation 1:

$$r_{PC} = \frac{n_{PC}}{n_{TOTAL}} \tag{1}$$

In this ranking approach, the sorting is still performed by the number of related molecules in the publications of an author. However, to distinguish between two authors with the same number of related molecules in their publications, we use the value of $r_{PC}$ which ensures that the author with a higher normalized publication count is ranked above other authors with smaller values of $r_{PC}$. Due to normalization, this ranking measure promotes authors who focus and publish more exclusively on the relevant related molecules and also promotes younger authors in the rankings.

Despite the differences between three ranking measures described above, all of them are based on counting the number of publications. Therefore, in the future we will refer to them as the publication count method while specifying a particular variation of the measure (hypergeometric, non-normalized, or normalized).

An alternative solution is to use a variant of the PageRank [34] algorithm. In the co-authorship network, the number of co-authored publications is determined by the strength of the collaboration. However, it might be possible for famous researchers who have a large number of publications over the years to occupy the "leader board" leaving young researchers who are relatively new to the field substantially lower chances of being recommended. Therefore, we propose an alternative definition of the strength of collaboration between co-authors, as given by Equation 2:

$$w = \frac{M_{x,y}}{\sqrt{N_x N_y}} \tag{2}$$

where $M_{x,y}$ is the number of co-authored publications of $x$ and $y$ on all molecules interacting with the molecule of interest, $N_x$ and $N_y$ are the total number of publications of $x$ and $y$ respectively.

In addition, to reduce the execution time of constructing the co-authorship network, for each individual molecule interacting with the molecule of interest we rank the pairs of collaborators by the number of the publications they co-authored. Compared to collecting the collaborators of all related molecules simultaneously, this divide-and-conquer approach iterates through the related molecules and finds the best collaborators with the most co-authored publications on each single molecule. Since the number of authors working on any given molecule is smaller than the number of authors working on all related molecules, this method ranks the best co-authors more efficiently.

### User Interface

User interface for the MoleClue application is implemented as a Web site. A solution based on the Web application is both lightweight in terms of resource requirements and provides a familiar environment to most users. All Web pages of the MoleClue application have a navigation bar at the top which allows users to perform a new search or navigate back to the main page. The search results can be sorted using either the PageRank or the publication count method. Regardless of the ranking method used for sorting the output, each search result entry contains the name of the author, a list of their publications on relevant molecules, and affiliation, if available. Additional technical information may also be provided, depending on the ranking method. For instance, for the PageRank algorithm, the raw value of the score and the number of neighboring molecules are shown. In order to ensure short response times and fast loading of pages, large result sets are not shown completely on one page. Instead, only the first portion (called a page) of the results is displayed. This mechanism, called paging, allows users to see the total number of result pages and navigate between them by operating designated controls located at the bottom of the page. Figure 3 shows the main page of the MoleClue Web user interface. When a user performs the search by the name of the molecule, the resulting list of authors looks similar to the one presented in Figure 4.

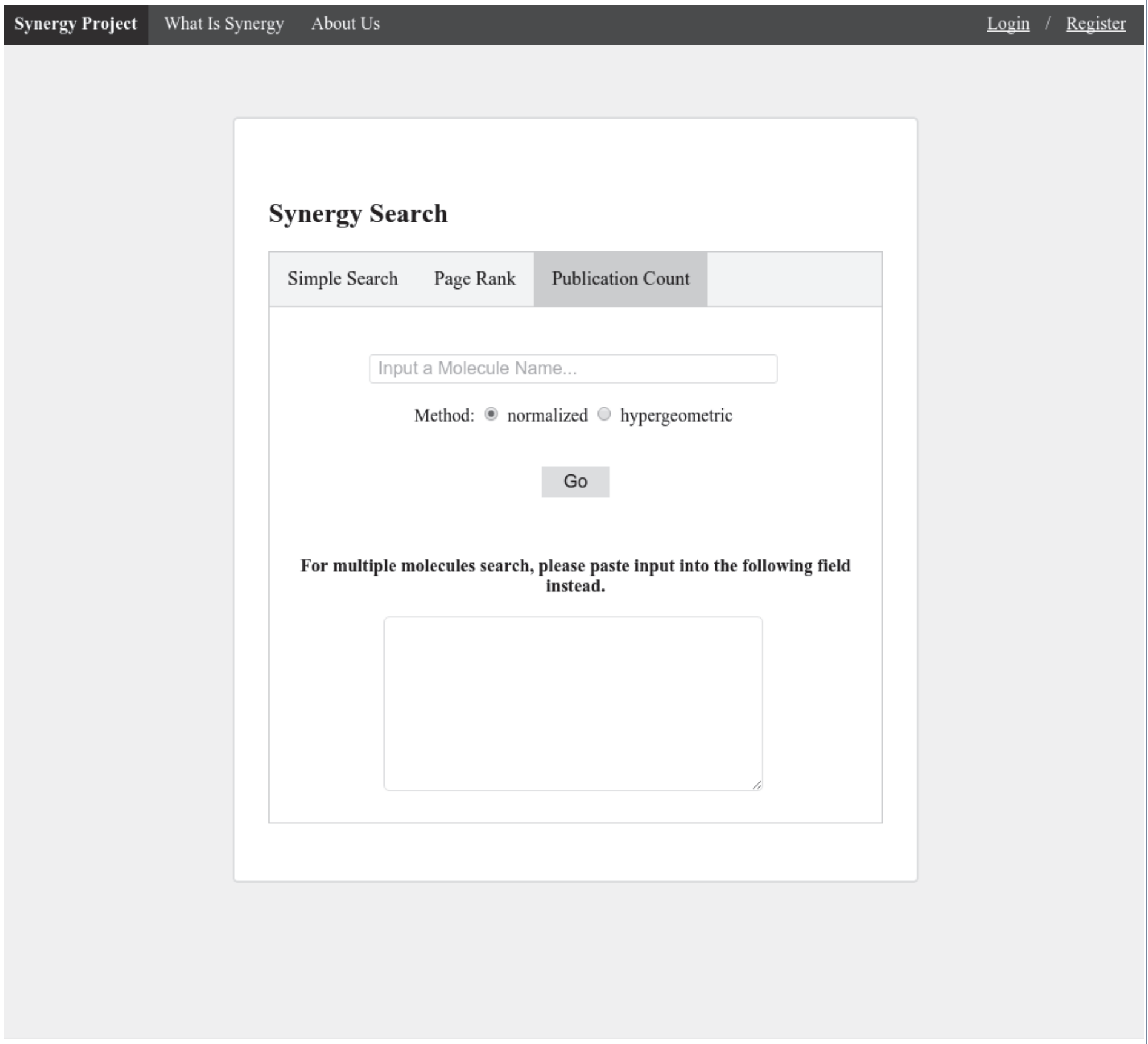

**Figure 3 MoleClue search page.** The name of the molecule should be entered into the corresponding text field. Pressing the Go button initiates the search.

## Experimental Results

### Testing Environments and Applications

Of all components of the MoleClue application, graph database storage demands the most computing power and is the most critical for the overall performance of queries. In our computing environment, an instance of the Neo4j 2.3.3 Community Edition is running under Linux operating system on a hyper threaded Silicon Mechanics Rackform iServ R420.v4 server. This computing platform has 32 cores structured as four Intel Xeon$^{TM}$ E5-4620v2 (2.6 GHz, 8-core, 20 MB Cache) processors and a 1 TB array of Random Access Memory (RAM) (32 x 32 GB DDR3-1600 ECC Registered 4R DIMMs) running at 1600 MT/s Max. Considering the actual load, each of the presentation and application layers is currently hosted on a dedicated standard desktop workstation. In the future, as the number of MoleClue users increases along with the load on the entire system, the Synergy Landscapes architecture allows the whole solution to be scaled both horizontally (by bringing online additional servers for the Web application layer and the graph database) and vertically (by adding resources to the existing nodes) to accommodate the growing demand.

All experiments were performed using the MoleClue application which has been created as a part of the Synergy Landscapes effort. We plan to make MoleClue available for use online free of charge; currently, demo results for the prototype implementation with a limited molecular network can be found at [42]. Additional materials for the entire Synergy Landscapes project and architecture are also provided online at [43].

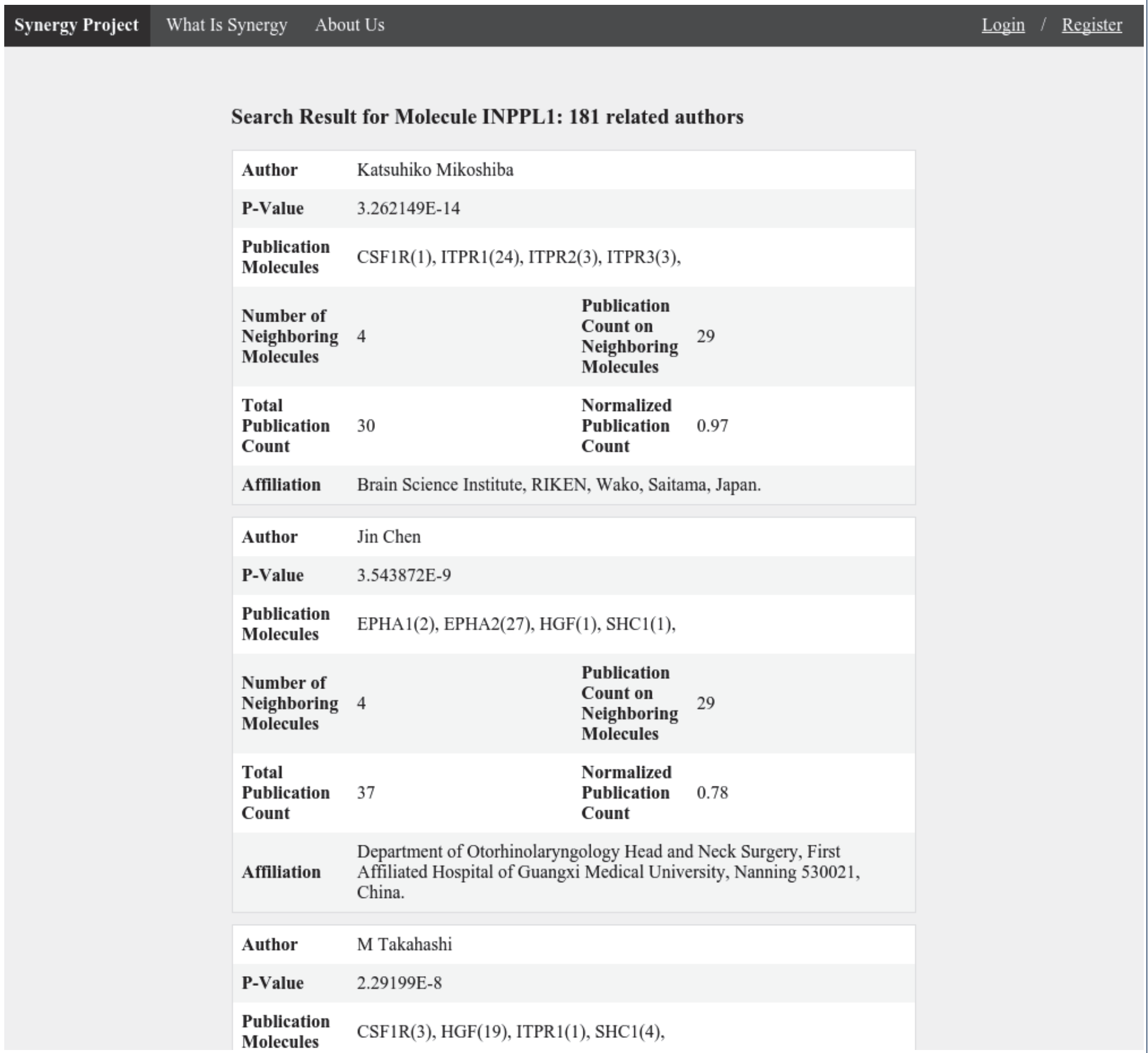

**Figure 4 MoleClue search results page for the publication count ranking.** Search results generated by MoleClue for the INPPL1 molecule. The list of suggested collaborators is ranked according to the publication count method and sorted by rank in the descending order.

The MoleCule application obtains its data from the PubMed [41] database. Publication records along with the author information are retrieved using the E-utilities API. These data are stored as nodes and edges in a local instance of the Neo4j graph database. An instance of the graph database which was used for performance evaluation contains data about 103 molecules, close to 64 thousand publications, and over 200,000 authors.

### Performance Evaluation

Considering the diversity of users' queries, specific co-authorship network is constructed according to the molecule of interest. Cypher queries are sent to the Neo4j database to selectively choose influential researchers working on the molecules related to the molecule of interest. The average time required to construct such co-authorship network is approximately 35.8 seconds and the corresponding variance is 5.95 seconds. In order to construct a precise co-authorship sub-network, the publications are filtered to eliminate those about irrelevant molecules dynamically. In this regard, such approach generally takes more time to response to queries but, as a benefit, it reduces the size of the co-authorship sub-network.

We illustrate the recommendation results produced by the PageRank algorithm using TREM2, INPPL1, and SORL1 as the molecules of interest. These molecules represent findings with novel relevance to the Alzheimer's disease. In order to improve our understanding of the role these molecules play in the development and progression of the disease, it might be beneficial to connect the related research.

Table 1 lists top five collaborators recommended for the TREM2 molecule along with their PageRank scores and the number of publications on the molecules interacting with TREM2. The data on top five collaborators for the TREM2 molecule according to the normalized version are presented in Table 2. The results suggest that by normalizing the strength of collaborations, the PageRank algorithm is likely to recommend the collaborators who have a relatively small number of publications compared to famous researchers.

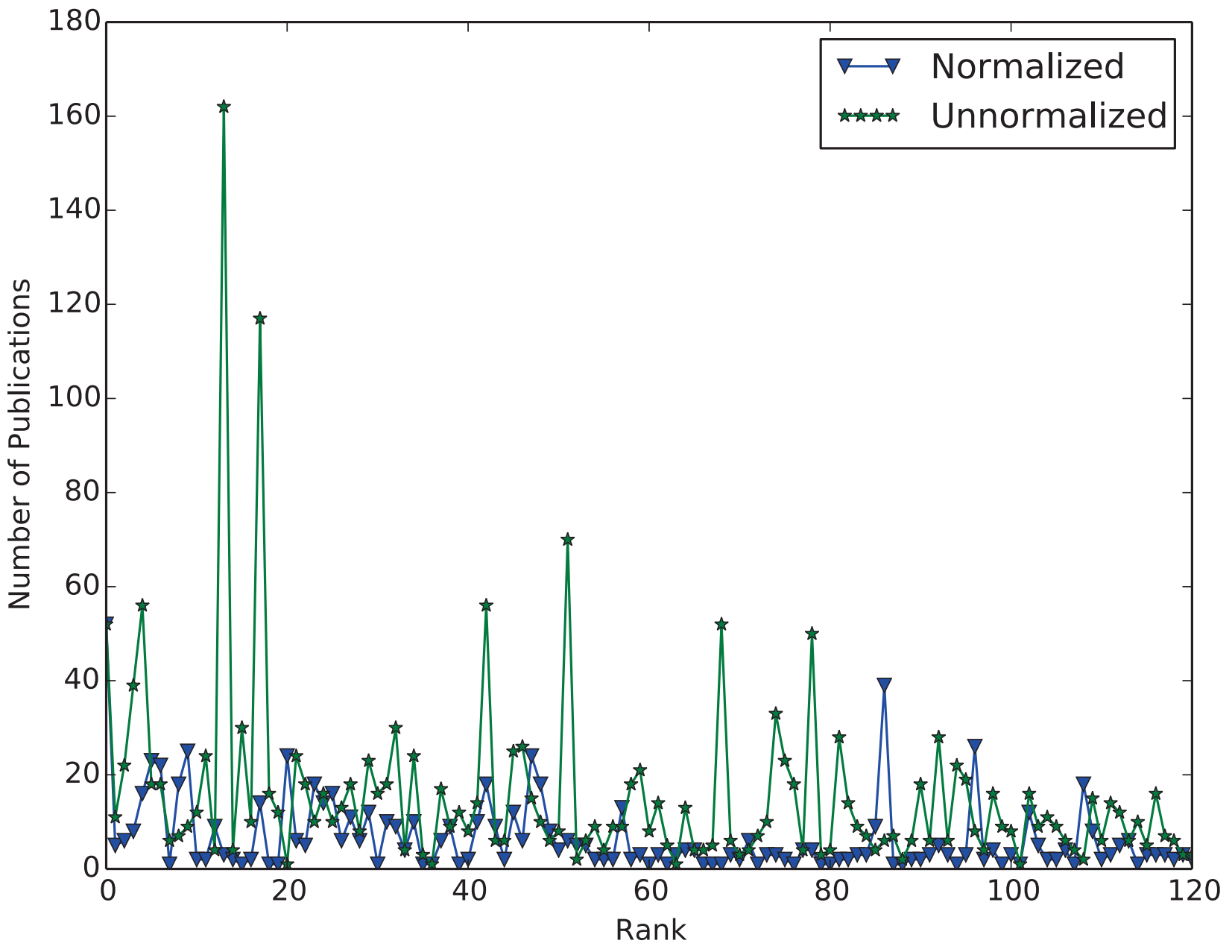

**Figure 5 Number of Publications vs. Collaborator Ranks in normalized and non-normalized versions.** Number of publications of the first 120 collaborators recommended for TREM2.

We compute the number of publications of the investigators who are ranked as the first 120 collaborators for molecule TREM2 and plot the number of publications vs. their ranks in Figure 5. It can be observed that the number of publications in the non-normalized version decreases more significantly than in the normalized version. Top five recommendations produced by the normalized PageRank method for the other two molecules which participated in our experiment (INPPL1 and SORL1) are given in Tables 3 and 4.

Apart from PageRank, earlier in part Network Analysis and Mining of the Synergy Landscapes Architecture section we discussed three ranking approaches to which we collectively refer as publication count ranking methods. The top five search results for these ranking methods based on the publication count (non-normalized, normalized, and hypergeometric) for the TREM2 molecule are given in Tables 5, 6, and 7. Ranking results for the other two molecules (INPPL1 and SORL1) and normalized and hypergeometric ranking methods are presented in Tables 8–9 and Tables 10–11, respectively.

**Table 1** Top 5 collaborators recommended for molecules related to TREM2 (non-normalized PageRank method).

| Name | Score | Related molecules (number of publications) |
|---|---|---|
| Marco Colonna | 0.00150 | ITGAM(8), VTCN1(2), TYROBP(42) |
| Holger K Eltzschig | 0.00096 | ADORA2B(19), ADORA3(3) |
| John S K Kauwe | 0.00079 | MAGI2(2), GRN(4) |
| Marina Cella | 0.00076 | ITGAM(4), TYROBP(12) |
| Carlos Cruchaga | 0.00075 | GRN(18) |

**Table 2** Top 5 collaborators recommended for molecules related to TREM2 (normalized PageRank method).

| Name | Score | Related molecules (number of publications) |
|---|---|---|
| Juha Paloneva | 0.00044 | TYROBP(5) |
| John S K Kauwe | 0.00040 | MAGI2(2), GRN(4) |
| Alberto Lleó | 0.00037 | GRN(8) |
| Marco Colonna | 0.00036 | ITGAM(8), VTCN1(2), TYROBP(42) |
| Sheng Chih Jin | 0.00035 | GRN(2) |

**Table 3** Top 5 collaborators recommended for molecules related to INPPL1 (normalized PageRank method).

| Name | Score | Related molecules (number of publications) |
|---|---|---|
| Anil K Sood | 0.00071 | EPHA2(40) |
| Nilufer Ertekin-Taner | 0.00060 | INPP5D(1), EPHA1(2) |
| Jin Chen | 0.00057 | SHC1(3), EPHA2(81), EPHA1(6), HGF(3) |
| Stephen Gottschalk | 0.00056 | EPHA2(4) |
| Robert L Coleman | 0.00051 | PIK3R1(1), EPHA2(10) |

**Table 4** Top 5 collaborators recommended for molecules related to SORL1 (normalized PageRank method).

| Name | Score | Related molecules (number of publications) |
|---|---|---|
| Marco Colonna | 0.00032 | PVRL1(15) |
| Ming-Yong Zhang | 0.00031 | ABR(1) |
| Masashi Narita | 0.00030 | ABR(1) |
| Josefina Casas | 0.00030 | ABR(1) |
| Jaime L Schneider | 0.00029 | ABR(1) |

As we discussed earlier, both normalized and non-normalized ranking methods perform initial sorting of the results based on the number of related molecules associated with the publications of a particular author. The data which we collected demonstrate that these methods produce different but close ranks for various authors, i.e., the effect of normalization is not very profound. In fact, for the SORL1 molecule, all five authors are the same as shown in Table 10, although their ranks are different. The hypergeometric test, however, produces strikingly different results. The top five produced by this ranking method are totally different from the top five produced by the other two methods, except for one author who is present in the top five with all the three ranking results for SORL1. The possible reason for this is that the hypergeometric test ranking does not account for the number of related molecules on which the author has published.

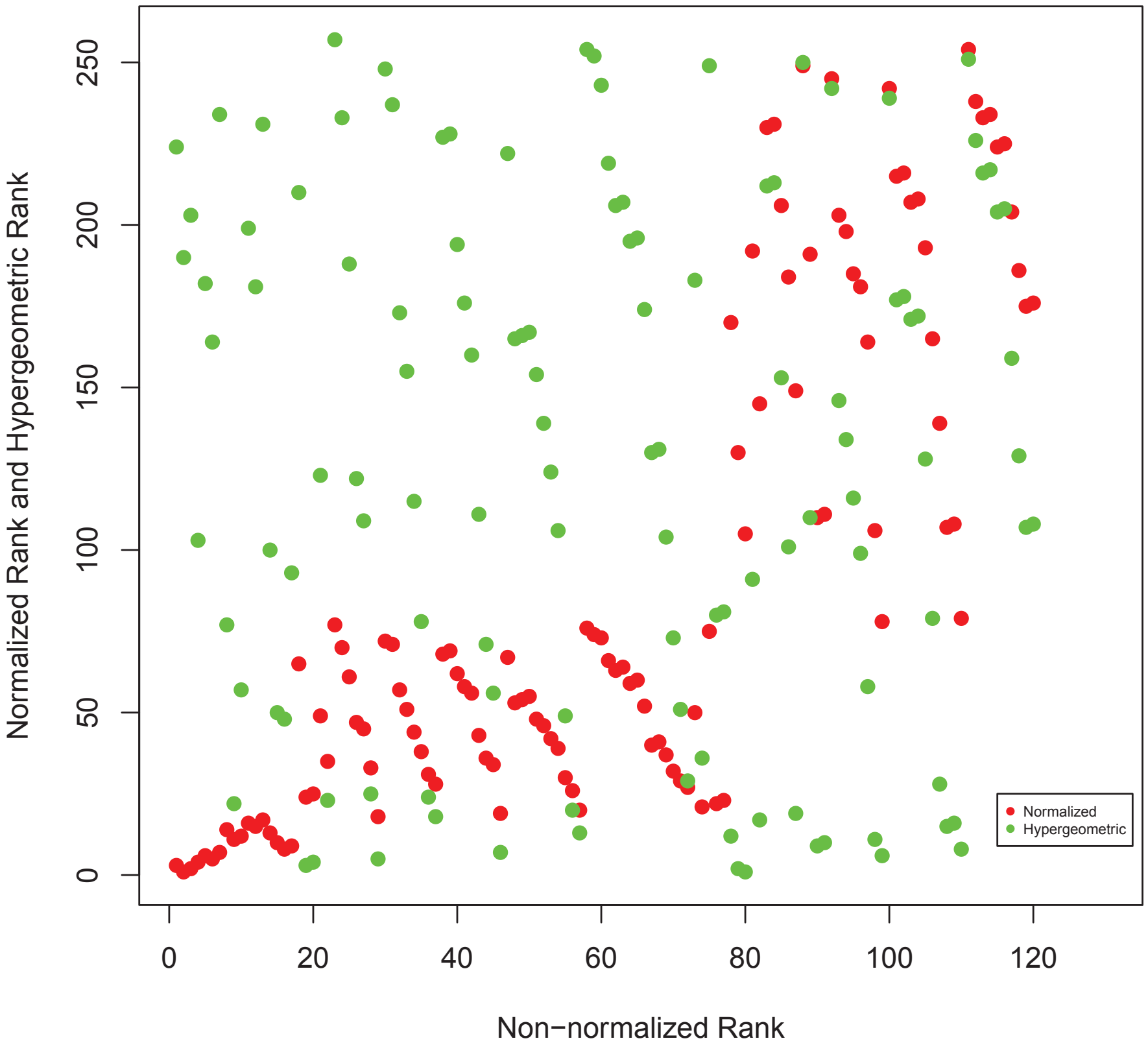

**Figure 6 Normalization of ranking results obtained with the publication count method for the TREM2 molecule.** First, a non-normalized version of the publication count ranking is applied to all authors who have at least one publication on TREM2 or any molecule associated with TREM2 in our network. Then, the top 120 names are selected from the resulting list. Finally, we separately apply normalization and the hypergeometric method to each author in the list and show how this affects the rankings.

**Table 5** Top 5 collaborators recommended for molecules related to TREM2 (non-normalized publication count method).

| Name | Related molecules (number of publications) | Number of related molecules | Publication count on related molecules |
|---|---|---|---|
| Wei Zhang | ADORA2B(1), ATF5(1), FPR3(1), GRN(2), ITGAM(6), TIMP1(1), TYROBP(2) | 7 | 14 |
| Ying Wang | FBP1(1), GPNMB(1), GRN(1), ITGAM(4), SLC1A3(1), SLC6A1(2), TIMP1(2) | 7 | 12 |
| Tao Wang | ACP5(1), CAPG(1), COLEC12(1), GRN(1), GPNMB(1), ITGAM(2), SLC1A3(1) | 7 | 8 |
| Li Zhang | ATF5(3), FBP1(1), GPNMB(2), ITGAM(7), SLC1A3(1), TIMP1(1) | 6 | 15 |
| Jing Li | FBP1(1), GRN(2), ITGAM(4), PLXNA1(1), TIMP1(3), TYROBP(1) | 6 | 11 |

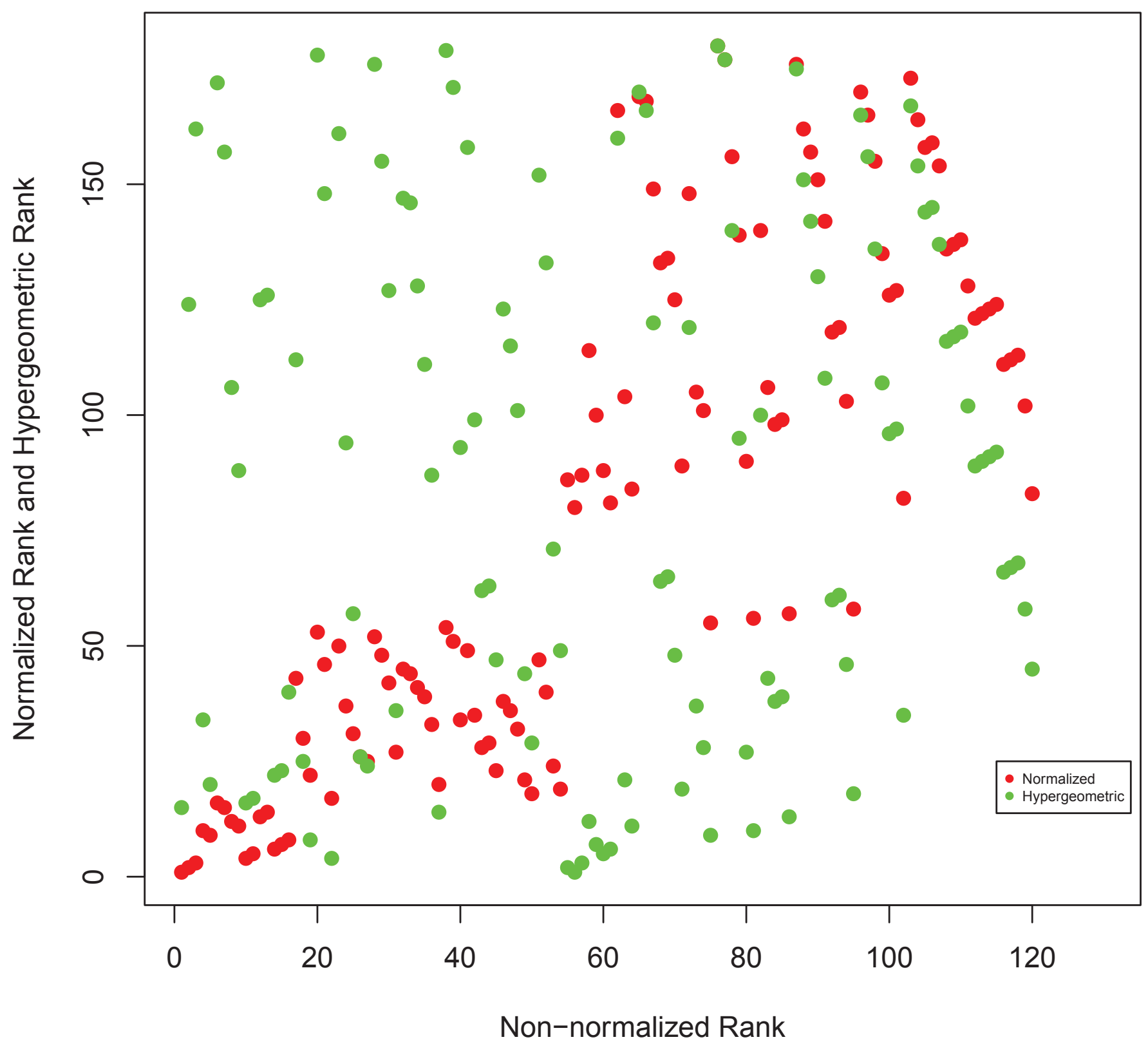

**Figure 7 Normalization of ranking results obtained with the publication count method for the INPPL1 molecule.** First, a non-normalized version of the publication count ranking is applied to all authors who have at least one publication on INPPL1 or any molecule associated with INPPL1 in our network. Then, the top 120 names are selected from the resulting list. Finally, we separately apply normalization and the hypergeometric method to each author in the list and show how this affects the rankings.

**Table 6** Top 5 collaborators recommended for molecules related to TREM2 (normalized publication count method).

| Name | Related molecules (number of publications) | Number of related molecules | Normalized publication count ($r_{PC}$) |
|---|---|---|---|
| Ying Wang | FBP1(1), GPNMB(1), GRN(1), ITGAM(4), SLC1A3(1), SLC6A1(2), TIMP1(2) | 7 | 0.21 |
| Tao Wang | ACP5(1), CAPG(1), COLEC12(1), GPNMB(1), GRN(1), ITGAM(2), SLC1A3(1) | 7 | 0.21 |
| Wei Zhang | ADORA2B(1), ATF5(1), FPR3(1), GRN(2), ITGAM(6), TIMP1(1), TYROBP(2) | 7 | 0.18 |
| Li Zhang | ATF5(3), FBP1(1), GPNMB(2), ITGAM(7), SLC1A3(1), TIMP1(1) | 6 | 0.28 |
| Wei Liu | GPNMB(3), ITGAM(3), MPDZ(2), SLC1A3(1), TIMP1(1), VTCN1(1) | 6 | 0.23 |

**Table 7** Top 5 collaborators recommended for molecules related to TREM2 (hypergeometric publication count method).

| Name | Related molecules (number of publications) | Hypergeometric $p$-value |
|---|---|---|
| Lieping Chen | GPNMB(1), ITGAM(4), VTCN1(16) | $6.51 \times 10^{-14}$ |
| Toshiyuki Takai | PLXNA1(1), SEMA6D(1), TYROBP(23) | $4.11 \times 10^{-11}$ |
| Lindsey A Criswell | GRN(1), ITGAM(10), MAGI2(1), VTCN1(1) | $2.39 \times 10^{-8}$ |
| Atsushi Kumanogoh | ITGAM(1), PLXNA1(9), SEMA6D(6), TYROBP(2) | $2.39 \times 10^{-8}$ |
| Noriko Takegahara | ITGAM(1), PLXNA1(6), SEMA6D(6), TYROBP(1) | $1.71 \times 10^{-6}$ |

**Table 8** Top 5 collaborators recommended for molecules related to INPPL1 (normalized publication count method).

| Name | Related molecules (number of publications) | Number of related molecules | Normalized publication count ($r_{PC}$) |
|---|---|---|---|
| Y Liu | CSF1R(5), HGF(22), HNRNPU(1), ITPR1(1), ITPR3(1), RPLP0(1), SHC1(2) | 7 | 0.53 |
| Yi Wang | CSF1R(4), EPHA1(2), EPHA2(1), GTF2I(1), HGF(2), HNRNPU(1), SHC1(1) | 7 | 0.36 |
| Lin Wang | EPHA2(1), GTF2I(1), HGF(3), ITPR3(1), PIK3R1(1), RPLP0(1), SHC1(1) | 7 | 0.24 |
| Piero Anversa | EPHA2(2), HGF(2), ITPR1(1), ITPR2(1), ITPR3(1), SHC1(2) | 6 | 1.00 |
| Annarosa Leri | EPHA2(2), HGF(2), ITPR1(1), ITPR2(1), ITPR3(1), SHC1(2) | 6 | 1.00 |

**Table 9** Top 5 collaborators recommended for molecules related to INPPL1 (hypergeometric publication count method).

| Name | Related molecules (number of publications) | Hypergeometric $p$-value |
|---|---|---|
| Katsuhiko Mikoshiba | CSF1R(1), ITPR1(24), ITPR2(3), ITPR3(3) | $3.26 \times 10^{-14}$ |
| Jin Chen | EPHA1(2), EPHA2(27), HGF(1), SHC1(1) | $3.54 \times 10^{-9}$ |
| M Takahashi | CSF1R(3), HGF(19), ITPR1(1), SHC1(4) | $2.29 \times 10^{-8}$ |
| Fumihiro Sanada | EPHA2(1), HGF(12), ITPR1(1), ITPR2(1), ITPR3(1) | $7.11 \times 10^{-8}$ |
| Y Watanabe | CSF1R(9), HGF(8), ITPR1(5), SHC1(1) | $3.17 \times 10^{-7}$ |

To further compare different publication count ranking approaches which we considered, the results of hypergeometric and normalized ranking methods have been plotted against the non-normalized ranking results. This has been done as follows: the top 120 authors have been identified in the non-normalized ranking scheme for each of the three searched molecules. For these 120 authors, their hypergeometric and normalized ranks have been noted. Then these two numbers have been plotted in Figures 6, 7, and 8 for TREM2, INPPL1, and SORL1 molecules, respectively. It shows that the non-normalized and normalized ranking schemes follow each other closely, while the hypergeometric ranks are different. This is supported by the cor-relation co-efficients between the rankings given in Table 12. We believe that the difference is caused by the fact that the hypergeometric publication count ranking method does not take into account the number of related molecules on which the author has published.

**Table 10** Top 5 collaborators recommended for molecules related to SORL1 (normalized publication count method).

| Name | Related molecules (number of publications) | Number of related molecules | Normalized publication count ($r_{PC}$) |
|---|---|---|---|
| Albert Hofman | MACF1(1), NISCH(1), PCSK7(1), PIK3CG(1), RERE(1), SORT1(2) | 6 | 0.47 |
| Jing Wang | ABR(2), BCL6(1), GGA1(1), GGA2(1), IQGAP1(1), PDLIM1(1) | 6 | 0.15 |
| Wei Zhang | BCL6(3), IQGAP1(1), MACF1(1), PIK3CG(1), PVRL1(1), SORCS1(1) | 6 | 0.10 |
| Eric Boerwinkle | MACF1(1), NISCH(1), PCSK7(2), PIK3CG(2), SORT1(4) | 5 | 0.67 |
| Christopher J O'Donnell | MACF1(1), PCSK7(2), PIK3CG(3), RERE(1), SORT1(3) | 5 | 0.62 |

**Table 11** Top 5 collaborators recommended for molecules related to SORL1 (hypergeometric publication count method).

| Name | Related molecules (number of publications) | Hypergeometric $p$-value |
|---|---|---|
| Nabil G Seidah | PCSK7(11), SORCS1(1), SORT1(1) | $1.55 \times 10^{-11}$ |
| Yong Liang | ABR(9), GGA1(1), IQGAP1(1), MACF1(1) | $1.07 \times 10^{-7}$ |
| Benjamin F Voight | MACF1(1), NISCH(1), PIK3CG(1), SORT1(4) | $1.51 \times 10^{-6}$ |
| Peder Madsen | GGA1(3), GGA2(2), SORCS1(3), SORT1(3) | $4.72 \times 10^{-6}$ |
| Eric Boerwinkle | MACF1(1), NISCH(1), PCSK7(2), PIK3CG(2), SORT1(4) | $7.09 \times 10^{-6}$ |

Based on the experimental data which we collected for three publication count ranking methods which we implemented and tested we conclude that in order to select authors who have published on a large number of related molecules and, thus, cover the neighborhood of the searched molecule, we should choose either non-normalized or normalized ranking methods. Out of them, the normalized ranking method promotes younger authors who are dedicated to research on the related molecules, while the non-normalized ranking method promotes more prolific, and possibly older authors. On the other hand, in order to select authors who have published more extensively and more exclusively on possibly a lower number of related molecules, we should choose the hypergeometric ranking method.

There is also little correlation between PageRank and three publication count measures. The possible reason is that PageRank identifies the most "influential" authors, while the publication count measures identify the most extensively-published and the most dedicated-to-the-related-molecules, i.e., prolific authors. The lack of correlation is clearly visible in Figures 9, 10, and 11, as most of the top 120 authors ranked by the non-normalized publication count method are absent in the top 120 list for both non-normalized and normalized versions of PageRank.

**Table 12** Pearson correlation coefficients for different publication count rankings.

| Variable | Value |
|---|---|
| $c_{TREM2}(Norm, Non-norm)$ | 0.81 |
| $c_{TREM2}(Hypergeo, Non-norm)$ | -0.05 |
| $c_{INPPL1}(Norm, Non-norm)$ | 0.81 |
| $c_{INPPL1}(Hypergeo, Non-norm)$ | 0.06 |
| $c_{SORL1}(Norm, Non-norm)$ | 0.58 |
| $c_{SORL1}(Hypergeo, Non-norm)$ | 0.01 |

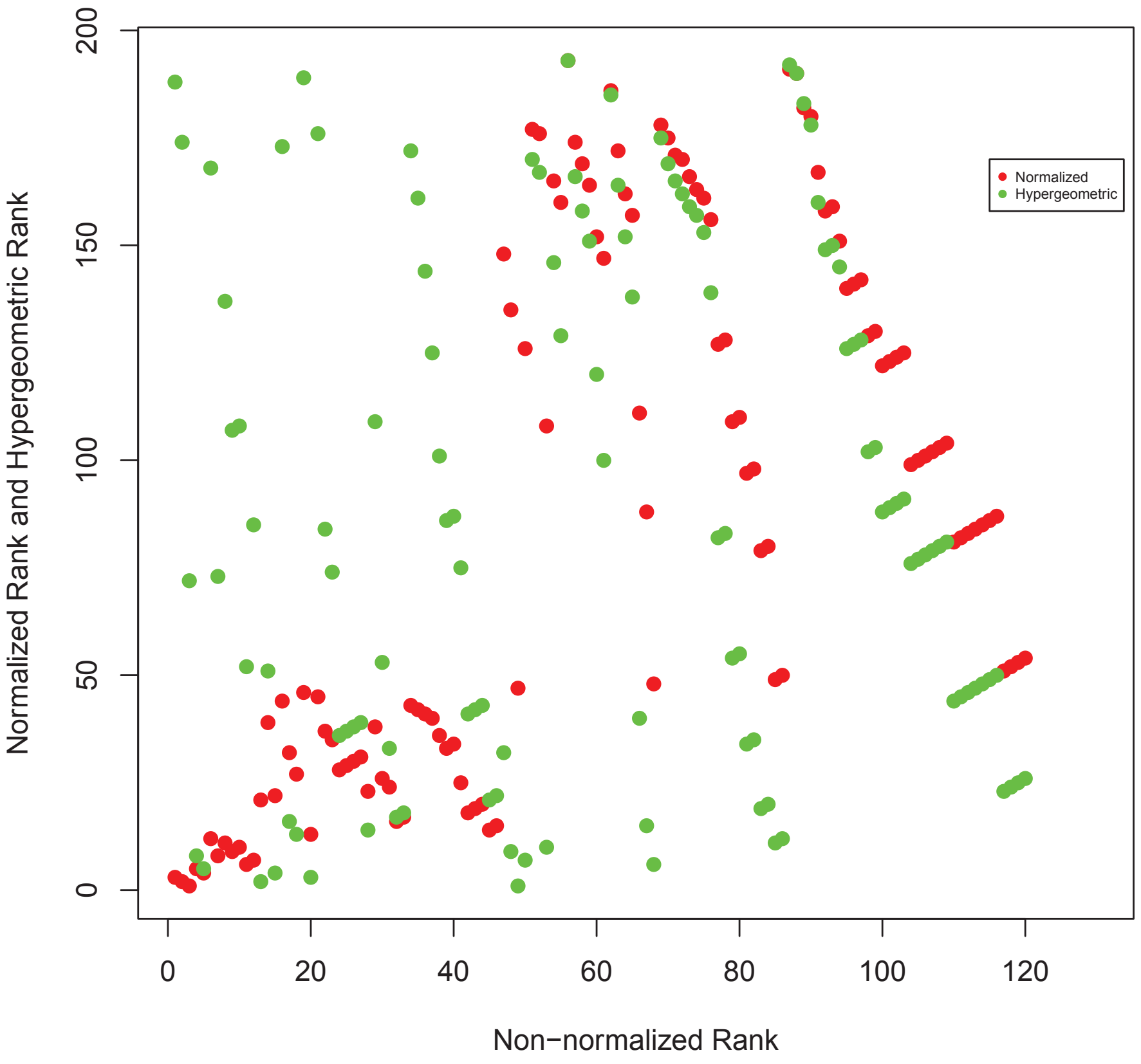

**Figure 8 Normalization of ranking results obtained with the publication count method for the SORL1 molecule.** First, a non-normalized version of the publication count ranking is applied to all authors who have at least one publication on SORL1 or any molecule associated with SORL1 in our network. Then, the top 120 names are selected from the resulting list. Finally, we separately apply normalization and the hypergeometric method to each author in the list and show how this affects the rankings.

In Figures 9 and 10, each dot $(x, y)$ represents a single researcher with rank $x$ and rank $y$ in associated approaches. It shows that the non-normalized version of PageRank gives results which are more similar to the results the publication count method gives than the ones obtained by the normalized version. As shown in Figure 10, the top 120 researchers recommended by the normalized PageRank algorithm are very different from the results produced by the publication count method, i.e. the two sets of top 120 results have a very small intersection. In contrast, the normalized and non-normalized versions of PageRank produce similar top 120 recommendations as shown in Figure 11. In addition, the order of top 120 recommendations depends on specific ranking schemes. In theory, identical top 120 recommendations should be presented as a straight line in Figures 9, 10, and 11. However, the observed results in these figures deviated significantly from the straight line shape, which suggests different orderings of the same group of researchers within the top 120 recommendations.

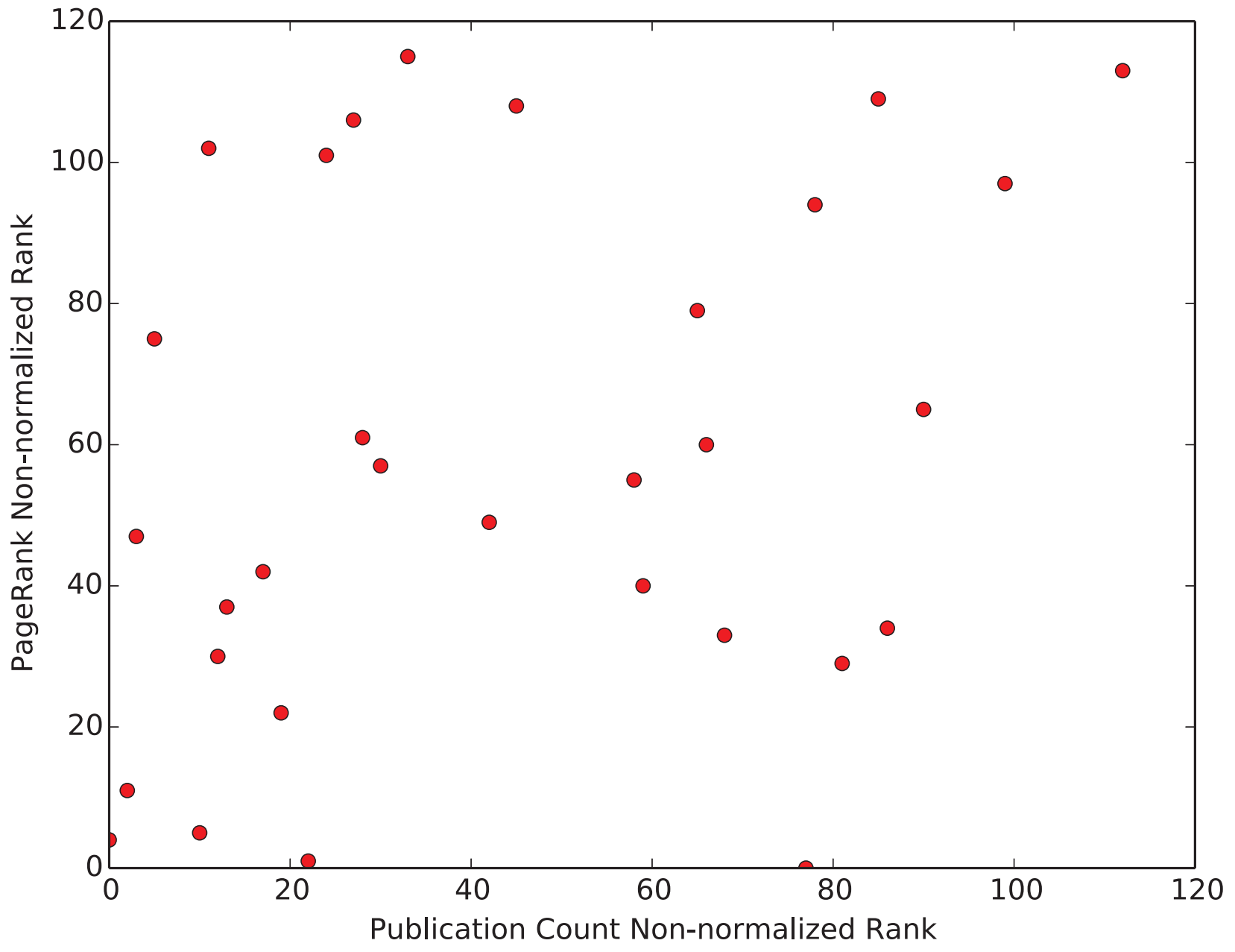

**Figure 9 Comparison of the top 120 results produced by PageRank (non-normalized version) and the publication count method on TREM2.** Ranks of the collaborators in top 120 recommendations produced by PageRank and the publication count method. Every dot represents a single researcher.

The execution times required to rank top 120 collaborators using different approaches are shown in Figure 12. In order to obtain the total number of publications on specific related molecules, the database has to traverse all paths from the particular author through their publications to the related molecules which takes a lot of time. In general, ranking researchers using the hypergeometric test and the other two publication count measures produces the results faster than the PageRank algorithm. In order to maintain the low latency response expected from an online system, intermediate results of the PageRank computation are stored as special service properties in the nodes. Subsequent invocations of the ranking procedure are capable of reusing those values instead of computing them from scratch. Using such precomputed attributes significantly speeds-up the perceived execution of user queries. In addition to that, the Web application layer is capable of caching user requests at the page level. In other words, when a user runs a query, the resultant HTML content which is sent back to the user's Web browser is also stored at the Web server for a certain period of time. When servicing the next request (from the same or other user), the Web server checks if cache contains the corresponding response. If so, then the request can be fulfilled directly from cache without the need to access the multilayer network data or initiate any computations.

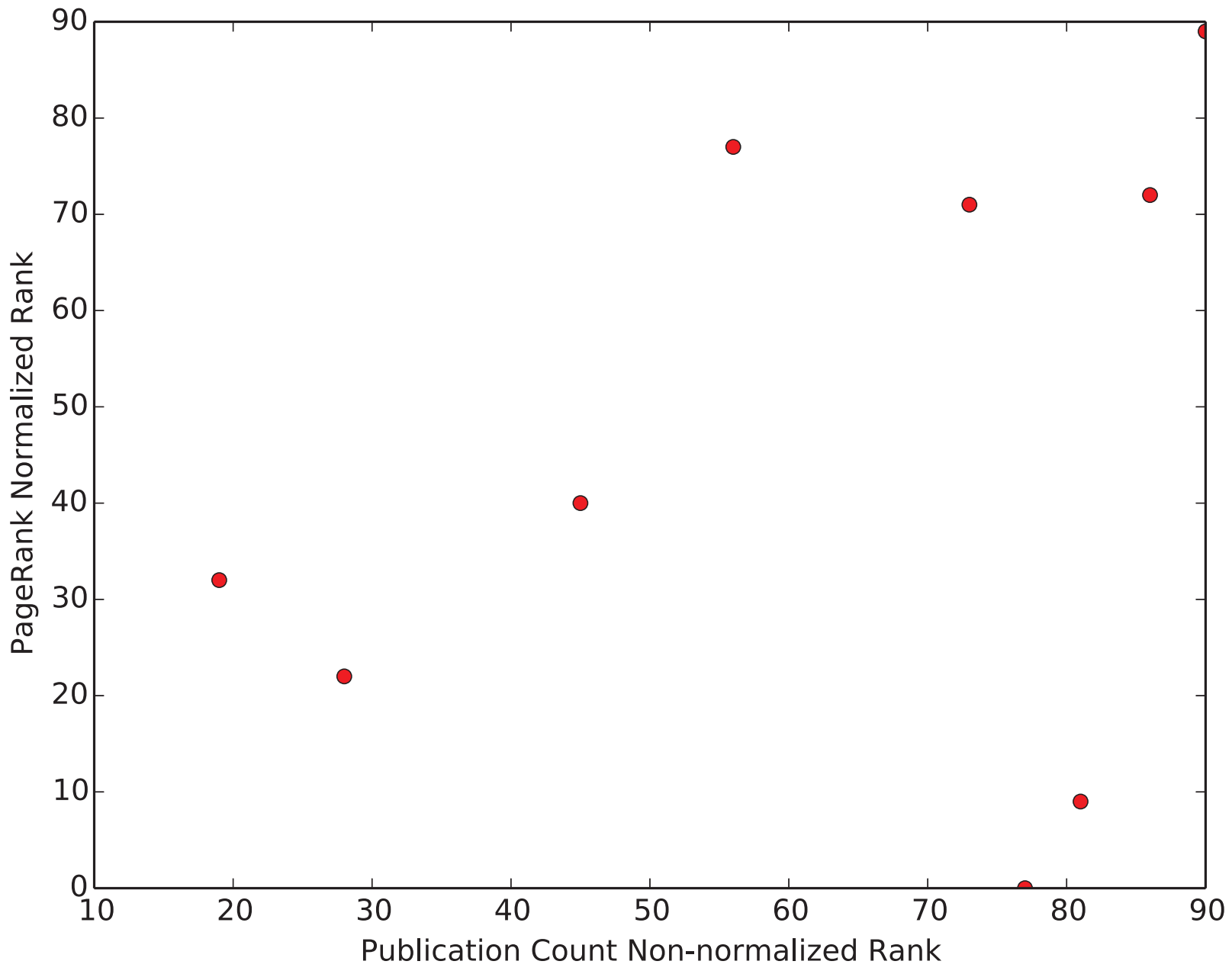

**Figure 10 Comparison of the top 120 results produced by PageRank (normalized version) and the publication count method on TREM2.** Ranks of the collaborators in top 120 recommendations produced by PageRank and the publication count method. Every dot represents a single researcher.

## Validation

The ultimate test of the utility of Synergy is the extent to which it increases scientific collaboration and high-impact interdisciplinary publications stemming from these collaborations. Since direct test of Synergy will require years to evaluate, it is important to find any limited evidence, available at the present time, that Synergy has the potential to increase efficiency and output of scientific collaboration. Currently, many factors influence selection of collaborators, such as university affiliation, grant opportunities, reputation, topical interest and methodological similarity. Among these and many other factors that influence the collaborator selection, the distance between topics of interests in molecular space has likely a minor influence - particularly when molecular proximity is based on recently-released omic datasets. Therefore, collaborations that conform to the Synergy model may be comparatively rare. However, if the underlying assumption of Synergy is correct - that scientists working on interacting molecules are more likely to engage in collaborative research - there may be some subtle evidence of this in existing publications.

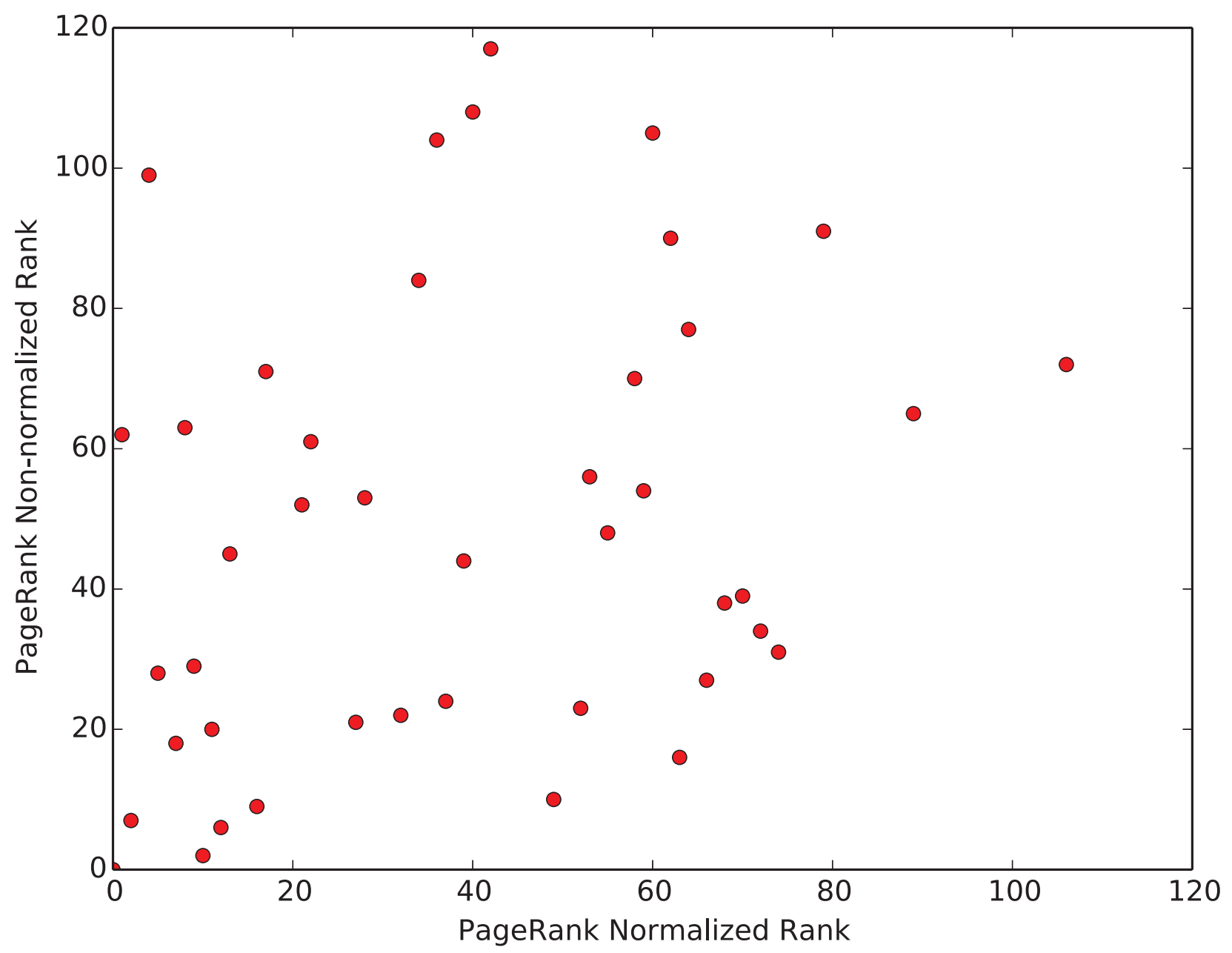

**Figure 11 Comparison of the top 120 results produced by the normalized and non-normalized versions of PageRank on TREM2.** Ranks of the collaborators in top 120 recommendations produced by the normalized and non-normalized versions of PageRank. Every dot represents a single researcher.

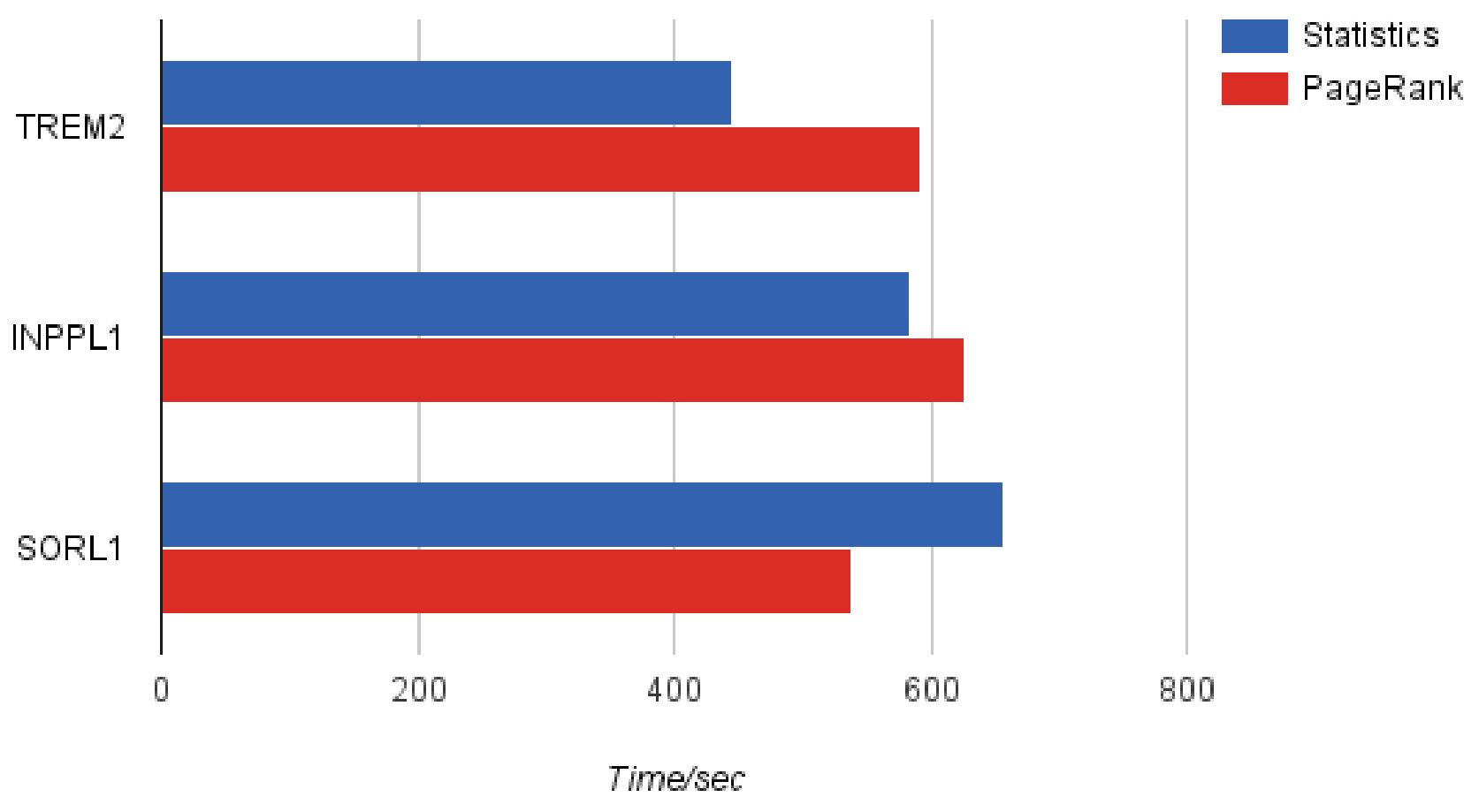

**Figure 12 Execution time of queries on TREM2, INPPL1, and SORL1 molecules (top 120 recommended collaborators).** The execution time required to obtain top 120 collaborators via ranking the statistics of researchers and PageRank algorithm.

To test if there is evidence that molecular interactions among scientists' topics of interest influence the likelihood of collaboration, we test if scientists who collaborate have molecular interests which are nearby in molecular networks. Specifically, we contrast the distance between molecules of interest to coauthors compared to randomly selected pairs of molecules. The molecular interests of a given author are defined as the 5 most commonly mentioned molecules in all of an author's abstracts. Collaborating authors are defined as an author's 5 most-frequent co-authors. For the null hypothesis, we select 500 random molecules, among which we find 10,330 pairs of neighbors, out of a possible 250,000 pairs (Table 13). We contrast this null with the proximity of molecules of interest to frequent coauthors. Specifically, we randomly select 250 authors with at least 5 publications. For each of these authors, we identify their top 5 collaborators, based on the number of coauthored publications. We identify the interests of each author and coauthor, defined as the 5 most frequently-mentioned molecules in their abstract. Then, after removing any molecules present the top 5 of both coauthors, we count the number of molecules which are network neighbors. Out of 668 pairs of molecules selected in this manner among coauthors, we find that 287 pairs are neighbors. The contingency table is given in Table 13 shows a significance in proximity of molecules of interest to frequent coauthors compared to random molecules ($p = 2.2 \times 10^{-16}$). The odds ratio for two molecules being neighbors when they stem from coauthors, compared to random selection, is 17.48. This indicates that molecular proximity influences likelihood of collaboration and supports the general concept of integrating molecular network proximity into collaboration prediction.

**Table 13** Validation Contingency Table.

| | Random Molecule Pairs | Author-CoAuthor Molecule Pairs |
|---|---|---|
| Non Neighbors | 239,670 | 381 |
| Neighbors | 10,330 | 287 |

## Conclusion and Future Work

The novelty and unexpected connections between topics that characterize award-winning scientific research [16] are also barriers to initiating such important projects, due to a competitive funding environment that values predictable results. Early-stage investigators may assist in introducing novel perspectives and experimental programs, but they are also the most vulnerable to risk, due to their career stage and lack of diverse funding streams. Therefore, in light of the conflicting public demands for high-impact advances, versus practical needs of scientists for "successful" research, we propose an efficient compromise, in the form of a tool that predicts innovative collaborations. This tool does not depend on a lengthy reform of scientific funding in order to achieve success, but rather attempts to identify situations where innovative findings are closely aligned with existing research projects. In this way, innovative ideas can be tested under the umbrella of existing programs, to provide crucial preliminary data in support of novel hypotheses.

The mechanism for detecting such innovative collaborations is aligning human interactions with cellular and biophysical interactions. To do this, we search through multilayer networks of molecular interactions and scientific papers, to identify paths that robustly link molecular topics of research. Because the accuracy and scope of of molecular interaction networks is increasing, and specialized molecular connectivity networks are generated for particular diseases and organisms, they provide a rational basis for predicting collaboration. At the same time, the massive structure of these networks means that no single human can account for all relationships between their work and other research programs. Therefore, it is helpful to employ a tool, such as Synergy Landscapes, that extracts predicted collaborations, from the multilayer network that links all of biological research. The ranking algorithm presented here is a proof of concept that it is possible to identify likely collaborators for a multilayer network, however, the exact ranking algorithm can likely be refined. As with results from any search engine, our results may be refined with user-controlled filters on geographic location or career stage, to increase their applicability to various situations.

Presently we search existing literature to identify researchers who are likely to be knowledgeable about particular molecular targets. However, the Synergy Landscape multilayer network could also be a clearing house for unpublished information, such as negative or preliminary results or unverified predictions from computational analyses. The molecular network assists in sharing these findings with people working in related areas, who might never search for these nearby molecules, particularly in the context of a disease outside of their specialty. In this way, the function of Synergy Landscapes is similar to craigslist, but in the context of molecular biology research. Proximity and local neighborhoods are defined through the structure of molecular networks, allowing users to exchange data and resources in a manner that is mutually beneficial. For instance, “abandoned” tool compounds developed in the context of a specific disease may be useful in the context of a different disease, whose molecular components are related. This ability to connect researchers may be particularly relevant for systems biology research, as it allows computational researchers to tag their top predictions in a public space, which is accessible to experimental scientists who may wish to validate predictions with a sufficient amount of evidence from computational analysis. All of these applications, which reuse or catalyze research depend on a tool to identify these possibilities. This demonstration shows that Synergy Landscapes could be a helpful tool in overcoming obstacles to innovative research.

**Competing interests**
The authors declare that they have no competing interests.

**Authors' contributions**
C. Gaiteri, K. Kuzmin, and B.K. Szymanski developed the high level concept of using mulitlayer collaboration network for nurturing innovative biomedical research. K. Kuzmin, X. Lu, P.S. Mukherjee, J. Zhuang, C. Gaiteri, and B.K. Szymanski designed the concept of study, developed the methodology, and performed the analysis and interpretation of data. K. Kuzmin wrote the code that collects the data from PubMed. X. Lu and P.S. Mukherjee wrote the code that performs data queries and implements different ranking methods. J. Zhuang implemented the Web application layer. K. Kuzmin, X. Lu, P.S. Mukherjee, J. Zhuang, and C. Gaiteri drafted the manuscript. K. Kuzmin, X. Lu, P.S. Mukherjee, J. Zhuang, C. Gaiteri, and B.K. Szymanski reviewed and revised the manuscript critically for important intellectual content and approved the version of the manuscript to be published.

**Funding**
Research was sponsored in part by the Army Research Laboratory under Cooperative Agreement Number W911NF-09-2-0053 (the ARL Network Science CTA), by the Office of Naval Research Grant No. N00014-15-1-2640, and by the Rush University grant to RPI. The views and conclusions contained in this document are those of the authors and should not be interpreted as representing the official policies, either expressed or implied, of the Army Research Laboratory or the U.S. Government.

**Availability of data and materials**

The Synergy Landscapes project uses the publication data from the PubMed [41] database which is developed and maintained by the National Center for Biotechnology Information at the U.S. National Library of Medicine located at the NIH. Access to PubMed is provided free of charge both through an online search capability and Entrez Programming Utilities [44] (E-utilities) which is one of the public Application Programming Interfaces (APIs) to all Entrez databases.

**Author details**

[1]Network Science and Technology Center, Rensselaer Polytechnic Institute (RPI), 110 Eighth Street, Troy, NY 12180, USA. [2]Rush University Medical Center, Rush University, 1653 W. Congress Parkway, Chicago, IL 60612, USA. [3]Społeczna Akademia Nauk, Łódź, Poland.